\documentstyle[12pt]{article}

\begin{document}
%\draft
%\preprint{}

\title{A fractal method for chaos in conservative
closed systems of several dimensions}

\author{ {\it A. E. Motter$^a$ and P. S. Letelier$^b$}\\
Departamento de Matem\'atica Aplicada - IMECC \\
Universidade Estadual de Campinas, Unicamp\\
13083-970 Campinas, SP, Brazil}

\date{}
\maketitle

\begin{abstract}

A fractal method to detect, locate and quantify chaos in
multi-dimensional-conservative-closed systems, based on
the creation of artificial exits, is presented. The method
is invariant under space-time changes of coordinates and
can be used to analyse both classical and relativistic
Hamiltonian systems of more than two degrees of freedom.
As an application of the method we study a couple of two
standard maps associated to a periodically kicked rotor
of $2\frac{1}{2}$ degrees of freedom.

\end{abstract}

\vskip 1.0truecm



$^a$E-mail: motter@ime.unicamp.br.

$^b$E-mail: letelier@ime.unicamp.br.

\newpage

\section{Introduction}

In the last century the mathematical basis of deterministic
chaos in compact phase space (closed) systems has become clear
\cite{guc}, and a number of methods have allowed the study of a
vast class of particular cases.
Important examples of these methods are
the Melnikov method, the Poincar\'e section method, and
the Lyapunov exponents method \cite{guc,ott}. Alas, the first
and second methods are hard to be used in high dimensional
systems and the Lyapunov exponents are coordinate dependent
\cite{matsas}. Accordingly, the majority of the
coordinate invariant results obtained so far refer to systems
of $1\frac{1}{2}$ or two degrees of freedom.

Fractal methods, on the other hand, are coordinate
independent and can be used in any dimension.
These methods have been largely applied in the
characterisation of the chaotic dynamics of dissipative systems,
especially in the study of attractors and
basin boundaries \cite{ott}. Fractal techniques
have also been employed to analyse scattering processes and
chaotic transients in conservative systems with exits (open systems)
\cite{bleher,chen}.

The aim of this Letter is to propose an invariant method
to detect, locate and quantify chaos in closed systems of
several dimensions. More specifically, we are interested in bounded
(recurrent)
motions taking place in the absence of attractors and {\it natural}
exits. The establishment of such a method is pertinent since the
standard fractal methods cannot be applied to these cases. Roughly
speaking, our method consists of
the definition of adequate {\it artificial} exits in the original
phase space, and the application of fractal type techniques to
analyse the exit orbits.  This {\it fractal method for closed
systems} (FMCS) provides a graphic means to locate the chaotic
and regular regions on phase space slices.
It also leads to an algorithm to determine the chaotic
and regular fractions of the phase space volume.
Different from the above mentioned methods,
the FMCS can be used to study classical as well as relativistic
Hamiltonian systems of more than two degrees of freedom.
In some sense the FMCS may be seen as a `higher-dimensional
generalisation of the Poincar\'e section method'.

First in this Letter we establish a conjecture that will support the
FMCS. In Section 3 we present the method divided in three parts:
($i$) to determine whether the system is chaotic or not;
($ii$) to locate the chaotic and regular regions on a two-dimensional
surface of section;
($iii$) to compute the chaotic and regular fractions of the phase
space volume.
In Section 4 we consider the application of the FMCS to a couple
of two standard maps.
Finally, we present our conclusions in the last Section.

\section{Chaos and fractals}

One could say that a system is chaotic if it presents
both sensitive
dependence on initial conditions and mixing in a nonzero volume
of the phase space. This definition, however, is not adequate
since it depends on the time parametrization of the system.
A time independent definition that solves this problem
was given by Churchill \cite{chur}: The system is chaotic if
it is topologically transitive and the set of
(points of) compact orbits is dense,
in some positive volume of the phase space. In less
technical terms, we define the system as chaotic
if it presents, besides transitivity \cite{deva},
a somewhere dense
set of infinitely many unstable periodic orbits.
Chaotic systems may exhibit not only chaotic but also nonchaotic
(regular) regions.
This concept of chaos is consistent with the general principle
that chaos prevents integrability in the chaotic regions.

We shall consider each $m$-dimensional
manifestly invariant part of the phase space
(e.g., each energy surface) of autonomous closed systems,
for which we define:
(1) An {\it exit} $E$ is an
$m$-dimensional region of the phase space, so that orbits
are considered out of the system when they arrive at $E$.
(2) The {\it attraction basin} $B_E$ is the
closure of the set of initial conditions whose orbits reach $E$.
(3) The {\it invariant set} $I_E$ is the set
of interior points of $B_E$ whose orbits do not reach $E$.

The possible chaotic behaviour of the system is
determined by the nature of its invariant sets.
For a small exit defined in a chaotic region\footnote{
The exit has to be sufficiently small
in order to avoid the complete outcome of
the invariant set.
},
the Hausdorff dimension \cite{ott} of the invariant
set is fractional and tends to the maximum value $m$
(the dimension of the ambient space) when the
size of the exit is arbitrarily reduced.
(In the case of nonhyperbolic dynamics, the maximum
value $m$ can be obtained for exits of finite
size \cite{lau}.)
On the other hand,
invariant sets associated to exits defined in regular
regions do not present any fractal
structure, and their dimensions jump (discontinuously)
avoiding fractional values when the exits are removed.
We conjecture that this behaviour is typical
for dynamical systems in general\footnote{
There are pathological examples of nonchaotic systems
that exhibit fractal properties
when exits are created (e.g., systems with degenerate
resonances on a Cantor set \cite{troll}).
It happens when the invariant set of the
exit system is fractal but nonchaotic.
This behaviour is, however, atypical.
}.

This conjecture states that chaos
in closed systems and fractals in exit systems
are both determined by unstable periodic orbits.
Fractal invariant sets are typically chaotic
in exit systems and chaos in closed systems
is associated to the existence of fractal
invariant sets. 
Therefore, the introduction and
removal of exits leads from one situation to the other.
Physically, an exit system with fractal invariant set
evolves chaotically for a period of time
before being scattered. When the exits are removed the
system evolves chaotically forever.
Further details about this conjecture can be found
in Ref. \cite{motter}.

\section{Fractal method}

The above conjecture can be used to obtain a method
to study chaos in conservative closed systems
of several dimensions
($2\frac{1}{2}$ or more degrees of freedom in the
Hamiltonian case).
In what follows we present one possible implementation
of such a method, the FMCS.

In order to have an insight about where to look for chaos,
we consider the volume of the phase space accessed by a
sample of orbits. The idea is that in a chaotic region
almost all orbits access approximately the same volume
(the volume of that chaotic component).
Regular orbits are expected to access lower dimensional surfaces
that are arranged in families (of torus in Hamiltonian systems)
with increasing area.

First we divide the phase space with a grid and then we
evolve a sample of random initial conditions for a large period
of time, counting the number of cells of the grid
visited by each orbit. 
In a histogram of the number of visited cells,
peaks suggest the possible existence of chaotic regions.
We refer to these regions as chaotic candidates.
With a search program, a subregion (a ball, for example)
can be located inside
each chaotic candidate. We define an exit in the subregion
of the chaotic candidate that we want to study,
and we compute (numerically) the dimension of the
corresponding invariant set.
Then we use our conjecture to conclude whether the
region is chaotic or not.
That is the first part of the FMCS.

Given an exit $E$ defined in a chaotic region, the chaotic
region itself corresponds to the attraction basin of $E$.
To locate the chaotic and
regular regions on a two-dimensional surface
$S$ (section) of the phase space, we define one exit in each
chaotic component of the system. Then we take initial conditions on
a grid in $S$ and we evolve these points until the 
rate of orbits arriving at the exits becomes negligible.
The regular regions, corresponding to points whose orbits do not
reach the exits, can be appropriately plotted in a two-dimensional
graph. In this graph, the chaotic regions are represented by the
blank area.
That is the second part of the FMCS.

A straightforward extension of the preceding paragraph's procedure
allows us to compute
the chaotic and regular fractions of the phase space volume. That is
the third part of the FMCS, where the chaotic regions are identified
from the evolution of points randomly chosen everywhere in the
phase space. The chaotic fraction ($c_f$) is then given by the
quotient between the number of initial points in the chaotic regions
and the total number of initial conditions. Naturally, the regular
fraction ($r_f$) is determined by the condition $r_f+c_f=1$. 

Finally, consistency tests can be made in order to check the
results and the hypotheses involved in the FMCS.
The possibility that two or more
disconnected chaotic components are associated to the same peak
of the histogram, for instance, can be verified by comparing the
fraction of the volume of each chaotic component with the
fraction of points in the corresponding peak of the histogram.
The volume of each chaotic region is
limited by the corresponding fraction of visited cells.
In addition, the stability of the results should be tested
by taking different grid sizes, periods of evolution,
exits, etc.

The last remark of this section concerns the computation
of the Hausdorff dimension of invariant sets.
A technique that demands a small computational effort
consists of computing the uncertainty dimension
\cite{lau} of the set of singularities of the
{\it escape time function}
defined on a smooth (one-dimensional) curve $C$:
the escape time $t=t(\lambda )$ is the time required
by the orbit of an initial point $x(\lambda )$ in $C$ to reach $E$,
where $\lambda$ denotes the curve parameter;
the singularities of $t$ correspond to points whose escape
time is infinite (points of the invariant set).
We measure the uncertainty dimension by applying
a statistical method presented in \cite{lau}.
A parameter value $\lambda_0$ is called $\varepsilon$-uncertain
if $\mid t(\lambda_0+\varepsilon)-
t(\lambda_0 -\varepsilon)\mid\geq \Delta$,
where $\Delta$ is a positive number.
We compute the fraction
$f(\varepsilon)$ of $\varepsilon$-uncertain
points for a large number
of random values of the parameter $\lambda$, and for
different values of $\varepsilon$ going to zero.
The asymptotic behaviour of this function is expected to be of
the form $f(\varepsilon)\approx\varepsilon^{(1-d)}$ for
any $\Delta$, where $d$ is
the uncertainty dimension. The nonfractal case
corresponds to $d =0$ and the fractal one to $0<d\leq 1$.
In the fractal case, the Hausdorff dimension of the (total)
invariant set is supposed to be typically equal to $D=m-1+d$.

\section{Example}

The KAM surfaces cannot isolate the chaotic layers in
nonintegrable
Hamiltonian systems of three or more degrees of freedom
\cite{lichtenberg}: All the chaotic volume is joined
together into a single global structure or, at most,
into a finite number of disjoint components.
In such components, chaotic orbits are expected to get
arbitrarily close to any point energetically accessible.
The hypotheses of our method are therefore satisfied
for each value of the energy, and the FMCS is supposed to work
in Hamiltonian systems of more than two degrees of freedom.
We follow with an example of a Hamiltonian-like
system that has an analytical expression for its discrete form.  

Section maps can be explicitly
obtained for periodically kicked rotors
of $2\frac{1}{2}$ degrees of freedom.
A simple example is given by the time dependent Hamiltonian
\begin{equation}
H=\frac{I_1^{\; 2}}{2 m_1}+\frac{I_2^{\; 2}}{2 m_2}
+\frac{f(\theta_1, \theta_2)}{(2\pi )^2}\sum_n\delta (t-n\tau ),
\end{equation}
where $\tau /m_1 =\tau /m_2 =1$ and
$f(\theta_1, \theta_2)=K\cos(2\pi\theta_1 )\cos(4\pi\theta_2 )$,
which is equivalent to an autonomous Hamiltonian of three degrees
of freedom for a fixed value of the energy.
Integrating the delta function at $t=n\tau$ we obtain a
four-dimensional map on
$[0,1] \times [0,1]\times [0,1]\times [0,1]$:
\begin{equation}
I_1^{(n+1)} =I_1^{(n)} +\frac{K}{2\pi}
\sin (2\pi\theta_1^{(n)} )
\cos(4\pi\theta_2^{(n)} ) \;\;\mbox{mod}\;\; 1 ,
\end{equation}
\begin{equation}
I_2^{(n+1)} =I_2^{(n)} +\frac{K}{\pi}
\cos(2\pi\theta_1^{(n)})
\sin(4\pi\theta_2^{(n)} ) \;\;\mbox{mod}\;\; 1 ,
\end{equation}
\begin{equation}
\theta_1^{(n+1)} =\theta_1^{(n)} 
+ I_1^{(n+1)}  \;\;\mbox{mod}\;\; 1 ,
\end{equation}
\begin{equation}
\theta_2^{(n+1)} =\theta_2^{(n)} 
+ I_2^{(n+1)} \;\;\mbox{mod}\;\; 1 .
\end{equation}
These expressions are coupled standard maps, and they
also represent a perturbed twist mapping \cite{lichtenberg}.

Let us consider system (2-5) for $K=0.5$.
We divide the phase space with a square grid of
$16\times 16\times 16\times 16$ ($65536$ cells).
Then we compute the number $N$ of cells visited by
the orbits of 150000
random initial points in 400000 iterations. The result
is plotted in the histograms of Fig. 1.
In Fig. 1a it can be seen that the histogram presents
only two main peaks. In Figs. 1b and 1c we show refinements
of each main peak. These peaks do not present smaller
isolated peaks. The peak near $N=0$ (peak I) is consistent
with the hierarchical distribution expected for regular regions.
The peak near $N=64000$ (peak II), however, may be associated to
chaotic regions.
We define as region I and region II the sets of initial
conditions associated to peak I and peak II, respectively.
Considering these initial conditions we find that all the
points in the ball of radius $r=0.23$ and centre
$(I_1,\theta_1,I_2,\theta_2)=( 0.278, 0.510, 0.450, 0.761)$
are in region II. We denote this ball by $E_0$,
and we use it as an exit to analyse region II.

In Fig. 2a we show the escape time corresponding to
points randomly taken on the curve
$C:=\{ I_1= \theta_1=\lambda, I_2=\theta_2=0.5\mid
0.1<\lambda<0.2\}$.
The complicated structure of this graph is present in arbitrarily
large magnifications of the interval. In Fig. 2b we show a $50$
times magnification that exhibits this property. That is a graphic
characteristic of fractal dimension. In fact, a numerical
computation of the uncertainty dimension
in the interval of Fig. 2a results in $d = 0.95\pm 0.01$ for
$10^{-13}<\varepsilon <10^{-9}$ ($\pm 0.01$ represents
the statistical error).
In the interval of Fig. 2b the dimension results in
$d = 0.97\pm 0.01$ for $10^{-13}<\varepsilon <10^{-9}$.
By numerically estimating the dimension in
smaller and smaller
intervals, we approach asymptotically to $d=1$.
This result is consistent with the nonhyperbolic
character of the system, since in nonhyperbolic
systems the invariant set dimension is supposed to be
maximal \cite{lau}.
Independently of the exact value of $d$,
the fact that $0<d\leq 1$ is
enough to conclude that region II is chaotic.

Now, let us consider the section
$S:=\{(I_1,\theta_1,I_2,\theta_2)\mid I_2=\theta_2=0.5 \}$.
In order to determine the regular and chaotic regions of $S$,
we compute the intersection with $S$ of
the attraction basin associated to the exit $E_0$.
For initial conditions taken on a grid
of $400\times 400$, with a good approximation,
the chaotic points
correspond to orbits that outcome in less than $50000$
iterations. Both chaotic and regular regions are
plotted in Fig. 3, from which some 
physical results can be directly obtained.
For example, since the chaotic region contains parts of
the lines $I_1=0$ and $I_1=1$, the periodicity in $I_1$
implies that the chaotic region actually runs from $I_1=-\infty$ to
$I_1=+\infty$ \cite{ott}. Thus, the evolution of an initial
condition in the chaotic region goes to arbitrarily large energies.
 
In addition, we compute the chaotic fraction of the
phase space volume. The attraction basin of $E_0$
is estimated by evolving 150000 random
initial points over 50000 iterations.
For $K=0.5$ the chaotic fraction
results in $c_f=0.975\pm 0.005$.
The confidence on this result remains in the fact
that no significant changes are observed for
100000 and 400000 iterations.
As a function of the parameter $K$, the chaotic fraction
goes from zero for $K=0$ (nonchaotic) to one for
$K\approx 1.1$ (completely chaotic). 

\section{Final remarks}

In this communication we presented a fractal method to
study chaos in closed systems of several dimension, the FMCS.
The method applies to conservative systems in general,
not necessarily Hamiltonian.
The FMCS is simple from the conceptual viewpoint
and is of easy numerical implementation, since it
consists of the adequate definition of exits and
the subsequent analysis of the corresponding attraction
basins and invariant sets.

The importance of this method is double:
(1) In classical systems,
that are provided by an absolute time parameter,
it appears as a systematic method to study chaos in
phase spaces of several dimensions;
(2) In relativistic
systems, that are invariant under space-time
diffeomeophisms, this invariant method presents the
nice property of avoiding coordinate effects,
i.e. the results are independent of the space-time
parameters used.

Finally, we consider the limitations of our method.
The first one concerns the difficulty of the method
in locating small chaotic regions. The method is more
efficient in systems with large chaotic regions.
In addition, the definition of
exits completely inside the
chaotic regions may be, sometimes, impossible.
The difficulty is due to the presence of small
regular regions embedded in the chaotic regions.
The effect of these small regions
is expected to be negligible, however.
The last problem refers to points near
the boundaries between chaotic and regular
regions. For these points, it is difficult
to determine whether they are
chaotic or regular.
All these problems are
intrinsic of numerical methods
and are also present, for instance, in the
Poincar\'e section method.

\vskip 0.5truecm

\leftline{ACKNOWLEDGMENTS}

The authors thank F. Bonjour, A.P.S. de Moura,
M.T. Ujevic and R.R.D. Vilela for their valuable
comments, as well as Fapesp and CNPq for financial support.

\newpage

\begin{figure}
\caption{Histograms of the number of cells visited in a
grid of $16\times 16\times 16\times 16$ for $K=0.5$.
The orbits of $150000$ random initial
conditions were computed over 400000 iterations.
(a) The histogram of all orbits.
(b) Refinement of peak I.
(c) Refinement of peak II.      
}
\label{figu1}
\end{figure}

\begin{figure}
\caption{Escape time relative to the exit $E_0$
for $K=0.5$.
(a) $10000$ points randomly taken on
$C:=\{ I_1=\theta_1=\lambda$,
$I_2=\theta_2=0.5\mid  0.1<\lambda<0.2\}$.
(b) A portion of (a) magnified 50 times.
}
\label{figu2}
\end{figure}

\begin{figure}
\caption{Portrait of the regular (in black)
and chaotic (in blank) regions on
$S:=\{(I_1,\theta_1,I_2,\theta_2)\mid I_2=\theta_2=0.5 \}$
for $K=0.5$. 
}
\label{figu3}
\end{figure}

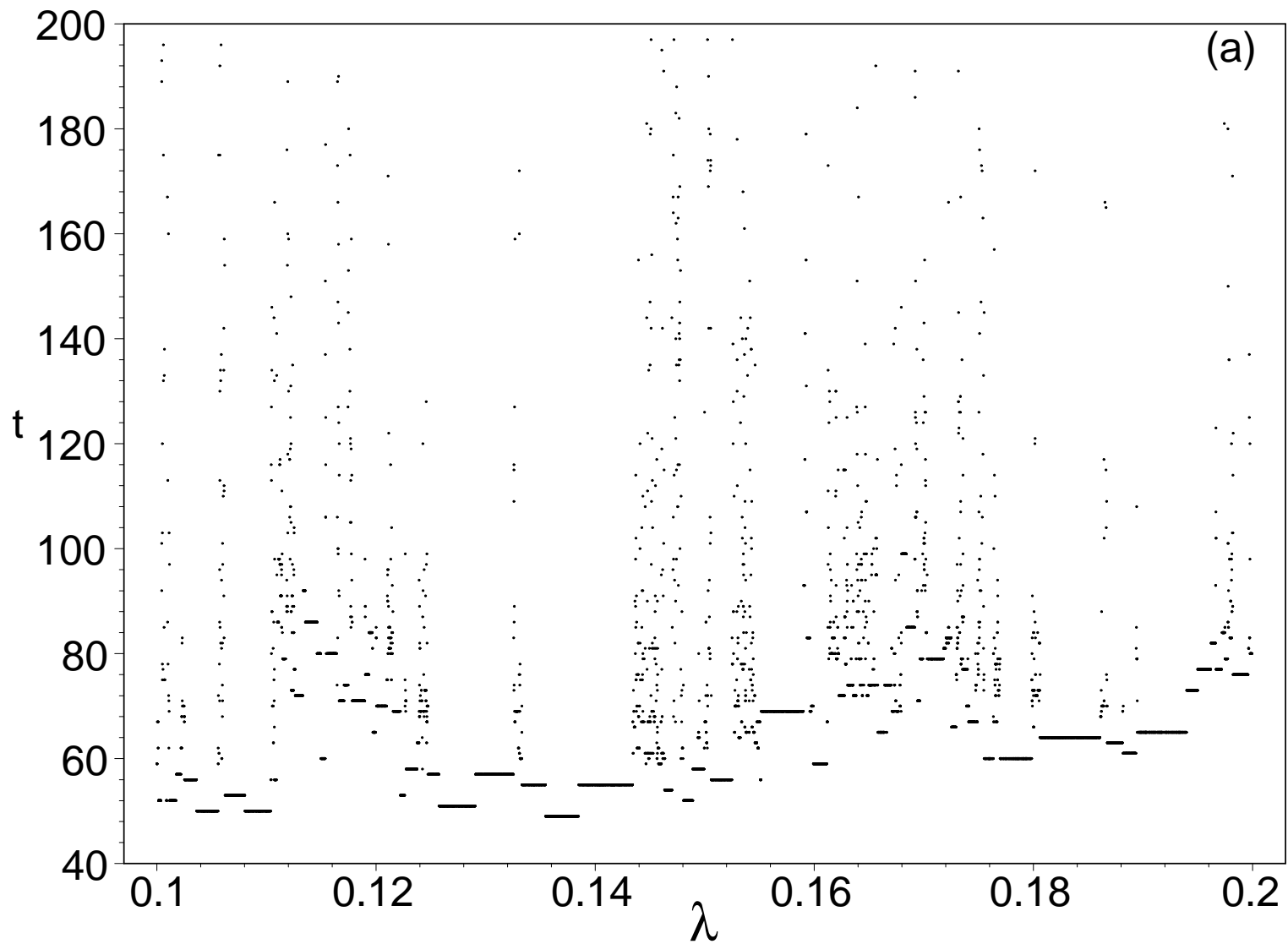
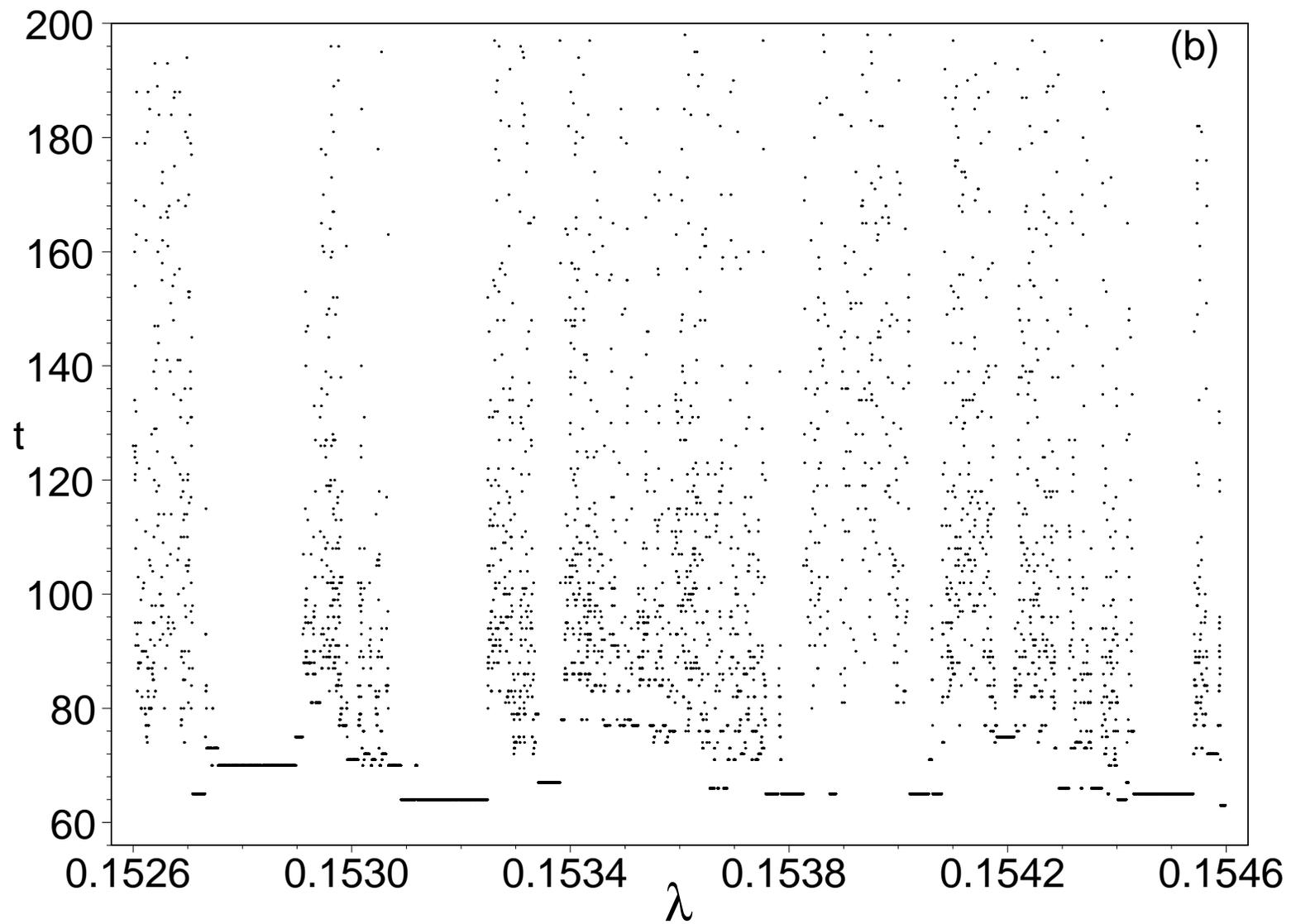
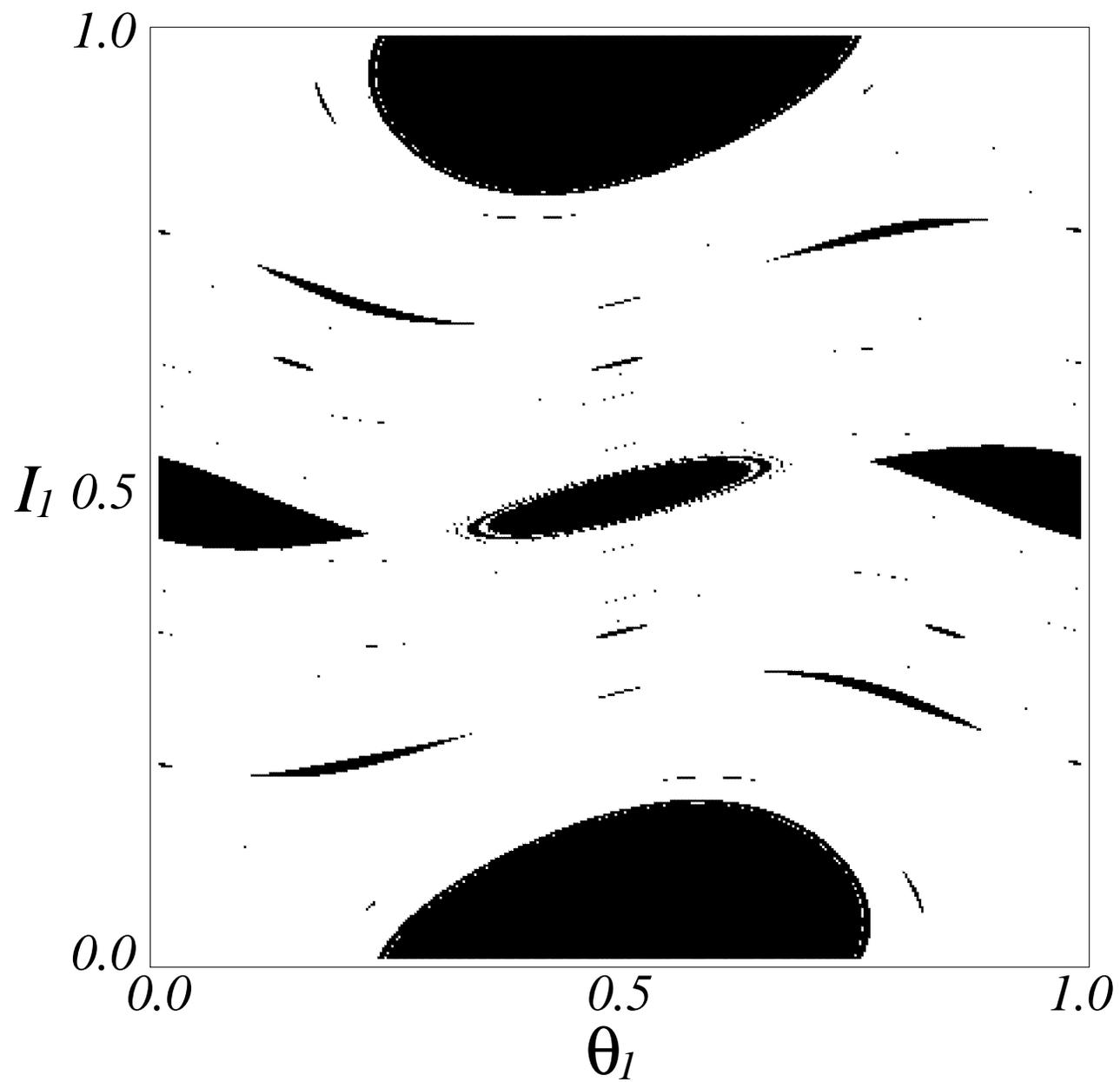
\end{document}